\documentclass[prl,superscriptaddress, amsfonts, amssymb, amsmath, reprint, showkeys, nofootinbib, twoside]{revtex4-2}
\usepackage[english]{babel}
\usepackage[utf8]{inputenc}
\usepackage{amsthm}
\usepackage{mathtools}
\usepackage{physics}
\usepackage{xcolor}
\usepackage{graphicx}
\usepackage[T1]{fontenc}
\usepackage[pdftitle={Article}, pdfauthor={Author}]{hyperref} % For hyperlinks in the PDF
\usepackage{physics}
\usepackage[colorinlistoftodos, color=green!40, prependcaption]{todonotes}
\usepackage{dsfont}
\usepackage{chemformula}
\usepackage{amsmath}
\usepackage{relsize}
\usepackage{enumitem}
\usepackage{epsfig}
\usepackage[amssymb]{SIunits}
\usepackage[normalem]{ulem}
\usepackage{esint}
\usepackage{multirow}
\usepackage{csquotes}

\renewcommand{\unit}[1]{\,\mathrm{#1}}
\addto\captionsenglish{}

\begin{document}
%\title{Electron-magnon scattering in complex systems from first principles} 
%\title{Magnon damping mechanisms in ultra thin metallic layers}
%\title{Frequencies and decay channels of terahertz magnons in ultrathin magnetic films: interplay of correlations, disoder, and multi-magnon processes}
\title{Correlations, disorder, and multi-magnon processes in terahertz spin dynamics of magnetic nanostructures: A first-principles investigation}
%\title{Frequencies and decay channels of terahertz magnons in ultrathin magnetic films: the role of correlations, disoder, and multi-magnon processes} 
%\author{Sebastian and the gang}
\author{Sebastian Paischer} \email{sebastian.paischer@jku.at} 
\affiliation{Institute for  Theoretical Physics, Johannes Kepler  University Linz, Altenberger  Stra{\ss}e 69, 4040 Linz} 
\affiliation{Department of Engineering and Computer Sciences, Hamburg  University of Applied Sciences, Berliner Tor 7, 20099 Hamburg, Germany}
\author{David Eilmsteiner} 
\affiliation{Institute for  Theoretical Physics, Johannes Kepler  University Linz, Altenberger  Stra{\ss}e 69, 4040 Linz} 
\affiliation{Department of Engineering and Computer Sciences, Hamburg  University of Applied Sciences, Berliner Tor 7, 20099 Hamburg, Germany}
\author{Igor Maznichenko} 
\affiliation{Department of Engineering and Computer Sciences, Hamburg  University of Applied Sciences, Berliner Tor 7, 20099 Hamburg, Germany} 
\author{Nadine Buczek} 
\affiliation{Department of Applied Natural Sciences, L\"ubeck  University of Applied Sciences, M\"onkhofer Weg 239, 23562 L\"ubeck,  Germany} 
\author{Khalil Zakeri} 
\affiliation{Heisenberg Spin-Dynamics Group, Physikalisches Institut, Karlsruhe Institute of Technology, Wolfgang-Gaede-Strasse 1, D-76131 Karlsruhe, Germany}
\author{Arthur Ernst}
\affiliation{Institute for Theoretical Physics, Johannes Kepler
  University Linz, Altenberger  Stra{\ss}e 69, 4040 Linz}
\affiliation{Max Planck Institute of Microstructure Physics, Weinberg
  2, D-06120 Halle, Germany}
\author{Pawe\l{} A. Buczek}
\affiliation{Department of Engineering and Computer Sciences, Hamburg  University of Applied Sciences, Berliner Tor 7, 20099 Hamburg, Germany} 
\date{\today}

\begin{abstract}
Understanding the impact of electronic correlations and disorder is essential for an accurate description of solids. Here,
we study the role of correlations, disorder, and multi-magnon processes
in THz spin dynamics. We reveal that a significant part of the
electron self-energy, which goes beyond the adiabatic local spin
density approximation, arises from the interaction between electrons
and a virtual magnon gas. This interaction leads to a substantial
modification of the exchange splitting and a renormalization of magnon
energies, in agreement with the experimental data. Finally, we establish a
quantitative hierarchy of magnon relaxation processes based on first
principles.   
\end{abstract}


\maketitle

Magnetic nanostructures and their intricate spin dynamics have fueled
remarkable developments in experimental, applied, and theoretical
quantum many-body physics over the recent years. On the applied side,
nanostructures, beginning with thin
films %\cite{Jourdan2015,Huang2018,Yang2021}
\cite{Yang2021} and proceeding down to single magnetic
atoms %\cite{Miyamachi2013,Natterer2017}
\cite{Natterer2017}, constitute the basis for spintronic and magnonic
information storage and processing
devices %\cite{Chumak2012,Gertz2015,Guo2019}
\cite{Guo2019}. A substantial body of research has been devoted to the
realization of quantum logical gates using magnetic degrees of
freedom %\cite{Chumak2015,Jia2021}
\cite{Jia2021}. Optical magnetization 
switching %\cite{Kimel2004,Kirilyuk2010,Guyader2015,Hadri2016}
allows to control the magnetization dynamics on the femtosecond timescale\cite{Hadri2016}.  For ultrathin magnetic films, recent experiments have yielded highly resolved spectra of the electronic (using, e.g., angle-resolved photoemission \cite{Sobota2021}) and, by means of spin-polarized high-resolution electron energy-loss spectroscopy (SPHREELS) \cite{Qin2019,Zakeri2021}, magnonic (spin-wave) band structures across the entire Brillouin zone.\\
On the other hand, the current first principles theoretical
description of spin dynamics lags clearly behind these spectacular
experimental developments. In this Letter we attempt to narrow this
gap. The task is of considerable interest, as the spectrum of
collective spin excitations (called spin-waves or magnons) determines
the thermodynamic properties of magnets, including the phase
transition temperatures. Additionally, the excitations contribute to
the specific heat as well as to the thermal and electric
conductivities. % \cite{Nolting2009}.
Furthermore, their coupling to electronic degrees of freedom can give
rise to a superconducting state \cite{Essenberger2016} and, in
general, influences the electronic band structure \cite{Tusche2018},
leading to a finite lifetime of excited electronic states
\cite{Schmidt2010a}. Last but not least, the damping of the spin
dynamics is of paramount practical importance in spintronic
applications %\cite{Krawczyk2014,Chumak2017}
\cite{Chumak2017} and, as we show here, constitutes an additional experimental probe sensitive to the atomic and electronic structure of nanostructures.\\
The density functional theory in a local spin density approximation
(LSDA) is able to provide a qualitatively correct picture of
electronic band structures of films and surfaces, including effects
like the formation of electronic surface and quantum well
states %\cite{Heinrichsmeier1993,Varykhalov2005}
\cite{Varykhalov2005}, but misses important corrections arising from
correlation effects \cite{Kotliar2006}, notably predicting a
substantially wrong value of the Stoner exchange splitting
\cite{Monastra2002}, among others.\\
In order to remain specific, we
  address Co films on different substrates for which a sufficient body
  of experimental evidence and theoretical studies concerning the electronic structure and spin
  dynamics is available
  \cite{Costa2004,Zakeri2021,Taroni2011}. Particularly, in the case of three monolayers (ML) of Co the fcc unit cell is complete and one expects to observe three magnon modes. As a matter of fact all these modes can unambiguously be resolved by the recent high resolution experiments by means of SPHREELS \cite{Chen2017,Zakeri2021}. 
 The latter SPHREELS experiments show that the theoretical description of these systems require a  substantial ad hoc \enquote{negative $U$ correction} of the occupied majority spin bands, as the LSDA drastically overestimates the spin-wave energies. Additionally, it falls short in accurately replicating the magnon lifetimes, performing well for Co/Cu but inadequately for Co/Ir.\\
 Different
descriptions of the band structure renormalization have been proposed,
including three-body scattering \cite{Monastra2002} and a sophisticated dynamical mean field
treatment \cite{Janas2022}. We have recently put forward a fully
\textit{ab initio} scheme \cite{Paischer2023} allowing to compute the electronic self-energy for complex systems, including two dimensional films, within Hedin's many-body perturbation scheme. This approach allows for the calculation of a non-local self energy while being numerically much cheaper than the aforementioned methods. Here, we show that in the considered Co films a
substantial part of these corrections arises indeed due to the
interaction of electrons with the gas of virtual magnons.\\ 
Likewise, the damping mechanism of THz spin excitations has not yet
been fully understood. While the Landau damping, arising due to the
interactions of collective spin-waves with single particle (Stoner)
excitations, is known to be an important decay channel in conducting
systems \cite{Buczek2011a, Qin2015}, it does not explain the entire
experimentally observed magnon peak width \cite{Zakeri2021},
especially in the high frequency range. Hence, other conceivable
damping mechanisms may arise due to spin dynamics occurring beyond the
linear response regime (expressible in the language of multi-magnon
processes \cite{Kaganov1987}), relativistic spin-orbit coupling (SOC)
\cite{Bergman2010,Bergqvist2013}, and the presence of structural
disorder \cite{Paischer2021a}. None of these effects have been systematically studied
within a realistic \textit{ab initio} framework so far.  Here, we
introduce a methodology capabale of accounting for all these
effects. Thus, for the first time
%Here we will provide a solution to all these problems and introduce a methodology which can account for all these effects. 
we provide a clear \textit{quantitative} hierarchy of spin dynamics damping mechanisms in itinerant magnetic films dominated by Landau mechanism and disorder with multi-magnon processes contributing weakly to the magnon linewidth.\\
%The text is organized as follows. First, the electronic ground state of the films is investigated taking into account many-body effects beyond the LSDA. Next, the spin dynamics is addressed within the framework of the linear response time-dependent density functional theory (TDDFT) in order to recover the energies and Landau damping of the magnon modes. Finally, we quantitatively describe the decay channels opening due to disorder and multi-magnon processes.
\textbf{Correlated ground state.} \\
%COUPLE OF SENTENCES ON THE METHOD? DYSON EQUATION? A PICTURE OF ELECTRON-MAGNON COLLISION?
The electronic structures are obtained using a first-principles Green's function method \cite{Hoffmann2020},
fully taking into account the effects associated with the
semi-infinite substrates and, if necessary, the disorder on the level
of the coherent potential approximation (CPA).\\
  For Co grown on
the Cu surface a major deficiency of the LSDA is the resulting
location of the occupied majority states too far below the Fermi
level. In turn, this yields a too large Stoner exchange splitting
between majority and minority bands. While the experimental value for
the exchange splitting for similar systems is reported to be around
0.8$\unit{eV}$ \cite{Miranda1983} the LSDA predicts values between
1.7$\unit{eV}$ and 2$\unit{eV}$ depending on the position in the
Brillouin zone. As shown below, this leads to substantial overestimations of
magnon energies. The origin of this shortcoming is that the LSDA
systematically fails to reproduce important correlation effects in the
band structure of magnetic 3d transition metals
\cite{SanchezBarriga2012}, influencing the values of the bandwidth and
exchange splitting, as well as the presence of satellite states. In
order to account for these effects, one must evaluate the electronic
self-energy, e.g. using Hedin's approach (many-body perturbation
theory (MBPT) framework) \cite{Nabok2021,Paischer2023}, evaluating
selected classes of Feynman diagrams. In particular, the possibility
of an electron or hole decay associated with the emission of a virtual
magnon accounting for the conservation of the spin angular momentum
(\enquote{electron-magnon interaction}), turns out to be essential in
the description of 3d magnets \cite{Paischer2023}. For bulk materials, the reduction of the exchange splitting has been shown to originate from the COHSEX self-energy \cite{Mueller2016} or the introduction of the electron-magnon interaction \cite{Mueller2019}. The latter effect has been also studied on the model level by Edwards and Hertz \cite{Hertz1973,Edwards1973}. While being state
of the art, such calculations are computationally demanding and so far
have hardly been applied to complex solids and nanostructures. In our
recently proposed computational scheme \cite{Paischer2023}, we
successfully approximate the corresponding series of ladder diagrams
in Hedin's theory \cite{Hedin1965} with less expensive response
functions and kernels available in the time-dependent density functional
theory \cite{Buczek2011a} allowing us to address systems as complex as
Co films considered here.
 \begin{figure}
 	\includegraphics[width=0.45\textwidth]{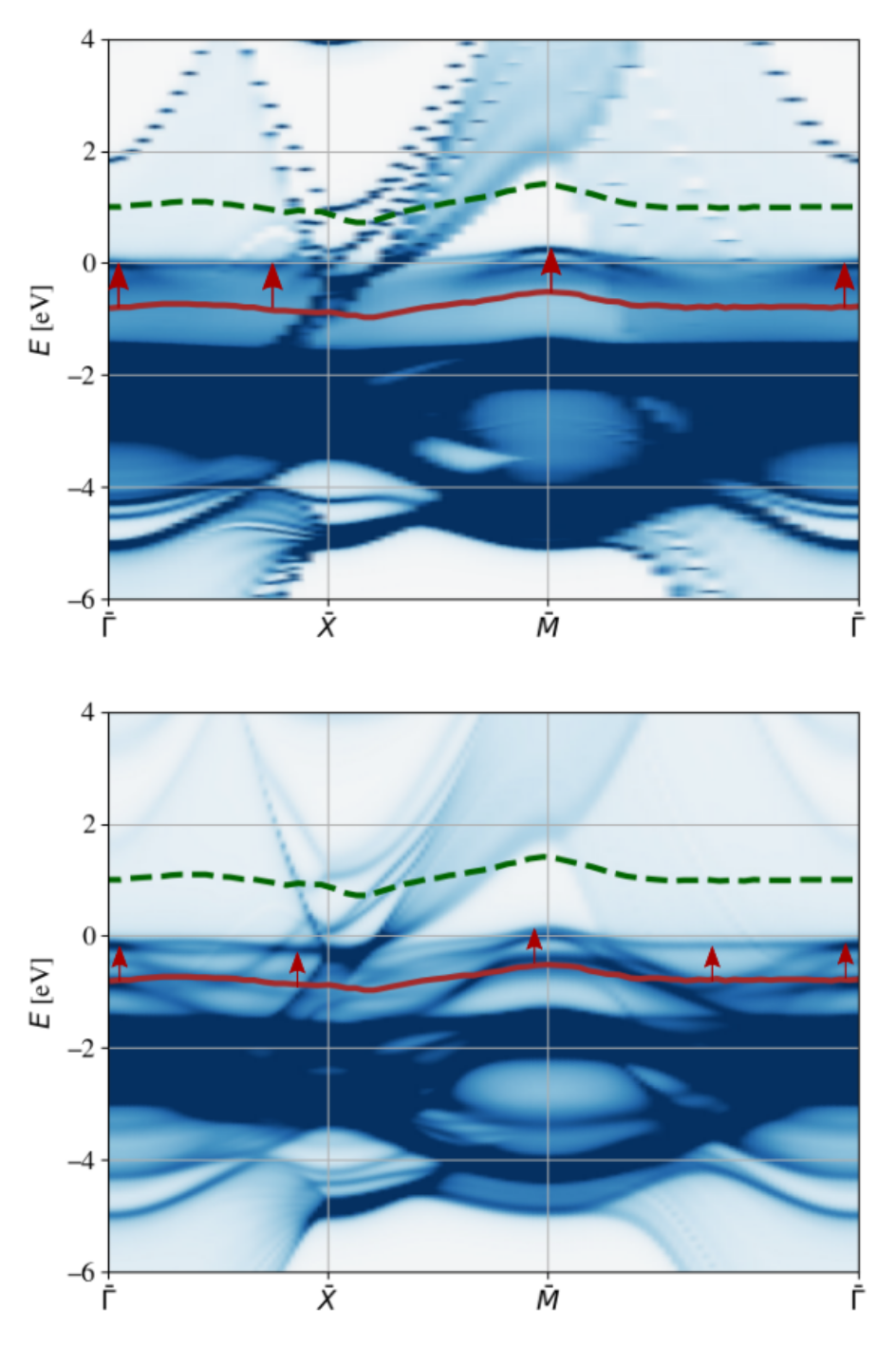}
 	\caption{Electronic spectrum of majority spin carriers in three layers Co on a Cu(100) substrate. %a) Result within the LSDA. b) Result including the electron-magnon scattering. c) Result for the disordered system within the LSDA+$U$ with $U=-1.6\unit{eV}$.
 	Top: Result within the LSDA including electron-magnon interaction (ordered system). Bottom: Result within the LSDA+$U$ with $U=-1.6\unit{eV}$ including disorder. The red/green solid/dashed line represents the highest almost horizontal band for majority/minority carriers within the LSDA. Arrows indicate the shift of the majority band through the electron-magnon iteraction (or the \enquote{negative $U$ correction}).}
 	\label{fig_3Co_Cu}
 \end{figure}
The upper panel in Fig. \ref{fig_3Co_Cu} shows the impact of spin-fluctuations on the
band structure of the 3$\unit{ML}$ Co/Cu(001) film. The
magnon-electron interaction shifts the occupied majority bands towards
the Fermi energy. The energy shift of the majority bands amounts to
approximately 0.8$\unit{eV}$, in agreement with results from hcp
cobalt \cite{Monastra2002,SanchezBarriga2012}. The quasiholes in the
majority spin channel acquire a finite lifetime being dressed now in
the gas of virtual magnons. The impact of the spin fluctuations on the
minority band is much weaker due to fewer electron partners in the
unoccupied spin-up channel for the exchange of the magnons. It turns out that the shift of the band
(but not the hole lifetimes) can indeed be modeled upon the
application of $U=-1.6\unit{eV}$ (in a standard LSDA+$U$ calculation) on the 3d bands of Co, cf. lower panel in Fig \ref{fig_3Co_Cu}. While being
methodologically limited, the latter approach allows nevertheless to
determine the band structure self-consistently which is still beyond
the computational reach of the current many-body approach (LSDA+$\mathcal{V}_{e-m}$). Consequently, we choose to use the LSDA+$U$ ground state as basis for the calculation of the magnon spectra as they are are sensitive to the details of the bandstructure close to the Fermi energy, where the self-consistency is particularly important.

Interestingly, in the Co/Cu system (but not Co/Ir), the LSDA+$U$
correction results in the ferromagnetic ground state becoming
unstable. At first glance this is not surprising, as the shift of the
bands towards the Fermi level results in long-range exchange
interactions between magnetic moments with oscillating sign. However,
this contradicts the experimental findings. This hints at a missing
element in the theoretical description.  According to the experimental
evidence \cite{Heinz2009,Nouvertne1999} both films feature a certain
degree of disorder, generally weaker for Co/Cu compared to Co/Ir. The
primary effect of disorder in Co/Cu is the smearing of majority bands
below the Fermi level as shown in the lower panel of Fig. \ref{fig_3Co_Cu}, the finite
lifetime arising from the collisions of electrons with the lattice
imperfections, and the consequent stabilization of the ferromagnetic ground
state. Nevertheless, the smearing is comparable with the electronic
lifetime acquired due to the exchange of virtual magnons. Thus, it is likely that both disorder and electron-magnon scattering are decisive
in the description of Co/Cu. However, as we shall show, it is crucial
in the description of the spin-wave \textit{damping} in Co/Ir.
\\
The impact of spin fluctuations on the band structure of the
3$\unit{ML}$ Co/Ir(001) film is significant as well, see supplementary
note II in \cite{supplement}. However, the impact of spin fluctuations cannot justify the value of $U = - 1.6\unit{eV}$ necessary for reproducing the experimental magnon
energies. The contrast between these two seemingly similar films
placed on the Cu and Ir substrates reveals the rich many-body physics
yet to be unveiled for nanostructures.

\textbf{Spin-wave energies and Landau damping.}

\begin{figure}
	\centering
	\includegraphics[width=0.45\textwidth]{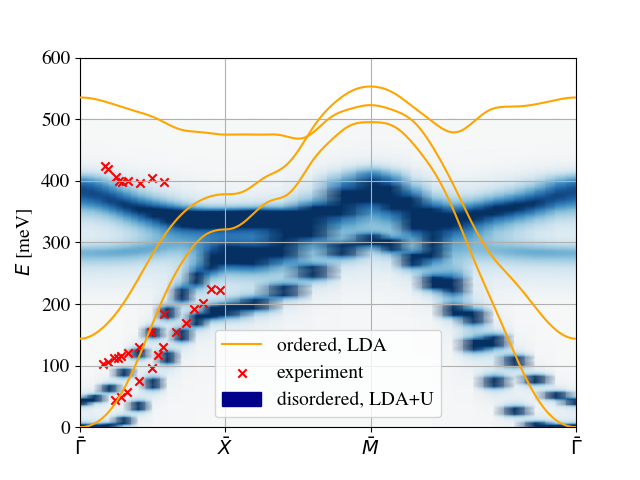}
	\caption{Magnonic spectrum of Co/Cu(100) calculated within the LSDA (orange lines) and with the LSDA+$U$ and disorder (blue background). The red crosses are experimental results \cite{Chen2017}.}
	\label{fig_Co_Cu_magnons}
\end{figure}

Figure \ref{fig_Co_Cu_magnons} shows the spin-wave dispersion for
Co/Cu calculated from a Heisenberg ferromagnet with exchange couplings obtained from the magnetic force theorem \cite{Liechtenstein1987}. %As mentioned above, the many-body corrections yield magnon energies in good agreement with experiment but taking into account the disorder is necessary to reproduce the stable FM ground state.
While the LSDA leads to magnon energies which are much too high (as
also discussed in \cite{Chen2017,Zakeri2021}), the inclusion of the \enquote{negative U correction} and disorder in the system leads to a
good agreement with the experimental results obtained by means of
SPHREELS \cite{Chen2017,Zakeri2021}. Note that the result for the
disordered system in Fig.\ \ref{fig_Co_Cu_magnons} has four magnon
modes while in the experiment only three modes were
observed. Theoretically, this is expected, as the magnetic Co atoms
are spread across four layers in the disordered system,
cf. supplementary note I in \cite{supplement}. The
spectral density of the almost dispersionless mode at
$E\approx300\unit{meV}$ is much smaller than for the other modes, which
we suspect to be the reason for its absence in the experimental
data. Closer inspection of Fig. \ref{fig_Co_Cu_magnons} reveals a very low spin stiffness. We believe that this is a remnant of the magnetic instability encountered when considering the ordered system within the LSDA+$U$. In the latter case the magnon energies are negative close to $\bar{\Gamma}$ as also shown in the supplementary note II  \cite{supplement}. The situation is similar for Co/Ir with a few exceptions. Most notably,
the ferromagnetic ground state is stable also after the
\enquote{negative $U$} correction. %In particular the system with 3 layers (with Co concentration of 40\% on the surface) had two high energy modes with very weak spectral weight.

Time dependent density functional theory (TDDFT) is capable of natively describing one of the dominating magnon
decay channels in metallic magnets, the Landau damping. It involves
the collision of the collective spin-wave with single particle
spin-flip excitations, the Stoner excitations. Our scheme \cite{Buczek2011a} involves the solution of the
\textit{susceptibility Dyson equation}
$ \chi^{\pm}=\chi_{\text{KS}}^{\pm}+ \chi_{\text{KS}}^{\pm}
K_{\text{xc}}\chi^{\pm}$ where $\chi_{\text{KS}}$ is the Kohn-Sham
susceptibility and $K_{\text{xc}}$ represents the exchange-correlation
kernel. The use of KKR Green’s function in the TDDFT calculations allows to describe the impact of the truly semi-infinite non-magnetic substrate on the spin-wave Landau damping. In the specific considered systems, the impact turns out to be small, due to the weak hybridization of substrate and film states. From the imaginary part of the susceptibility we extract the width of the magnon peaks utilizing a Lorentzian fitting function (the same method was used for disorder induced damping discussed later). The Landau damping is believed to be
dominating in conducting nanostructures (except for half-metals
\cite{Buczek2009}) and reproduces the SPHREELS data for Co/Cu
\cite{Zakeri2021}.
%For Co/Cu, the theoretical data for the Landau damping even slightly overestimates the magnon linewidth, as reported in \cite{Zakeri2021}, which might be attributed to the failure of the LSDA+$U$ to account for the damping of electronic states. (ICH VERSTEHE DIESEN LETZTEN SATZ NICHT.)
However, in the case of Co/Ir(001), the damping rate for optical
terahertz magnonic bands is clearly underestimated
\cite{Zakeri2021}. This hints at another important spin-wave
attenuation mechanism operative in ultrathin magnetic films. \\
The comparison of experimental results, in particular the magnon damping, obtained at room temperature and at 10K showed nearly the same results, indicating the small role of temperature induced magnon decay in this temperature range. This is not surprising, as the Curie transition in the films considered here occurs far above the room temperature \cite{Schneider1990} and at significantly lower temperatures there are not many thermally excited magnons which could contribute to the decay processes. Correspondingly, our theory targets the low temperature regime.
%It should be mentioned at this point that the values for the Landau
%damping are calculated for an ordered system. 

\begin{figure}
	\centering
	\includegraphics[width=0.45\textwidth]{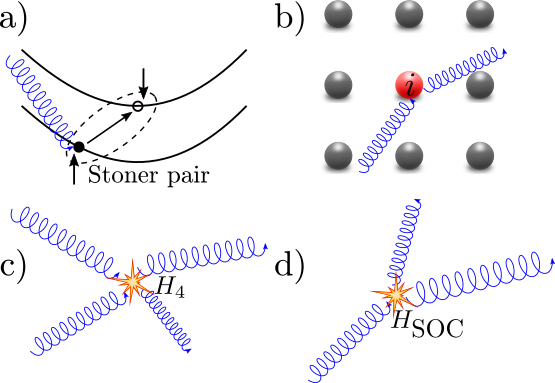}
	\caption{Magnon (depicted as blue arrows) damping mechanisms in itinerant electron ferromagnets. a) Landau damping: The magnon is absorbed by an electron which is excited in the process. b) damping through disorder: The magnon scatters on a crystal impurity $i$. c) Four-magnon processes: Two magnons scatter on each other. d) Three-magnon processes: One magnon decays into two magnons.}
	\label{fig_damping_channels}
\end{figure}

\textbf{Non-Landau magnon decay channels.}\\
Only few studies on other than Landau magnon damping have been published thus far. For instance, some research has been conducted at an analytical level regarding scattering on impurities \cite{Arias1999,McMichael2004,Zakeri2007}. 
%Only little attention has been paid in the literature to other than Landau-like decay channels of spin-waves in ultra-thin magnetic films. 
Here, we quantify them in an \textit{ab initio}
scheme. Conceivable non-Landau damping channels are depicted
schematically in Fig. \ref{fig_damping_channels} and discussed in
detail in the following.

Let us consider magnon-magnon interaction contribution to the lifetime
first. We discuss the process on the level of the Heisenberg
Hamiltonian. In general, a magnon can decay into one or more magnons
\cite{Kaganov1987}. However, in the absence of the spin-orbit coupling
(SOC), in ferromagnets, the magnons are elementary excitations and
eigenstates of the magnetic system. When there is no excited gas of
spin-wave bosons to interact with, this channel is inactive. The
observation pertains to the TDDFT as well. In half-metals, under weak
SOC assumption, there is no Landau damping and the spin-waves do not
decay \cite{Buczek2009}. An excitation of multiple magnons in a
certain process corresponds to large precession amplitudes of magnetic
moments and is a non-linear effect which cannot be grasped in the
linear response theory. However, the SPHREELS involves a continuous
generation of magnons caused by the electron bombardment of the
sample. Thus, even at low temperatures of the experiment when there
are no thermally excited spin-waves, a magnon gas can form,
facilitating the decay. A direct magnon-to-magnon decay cannot occur
(the magnons being eigenstates) but a given magnon coupled with
another magnon of the gas can decay into new pair of magnons
(\enquote{four-magnon-process}). With SOC present, a magnon can
furthermore decay into two magnon states with the lattice absorbing
the excess angular momentum (\enquote{SOC-} or
\enquote{three-magnon-process}).

\begin{figure}
	\centering
	\includegraphics[width=0.45\textwidth]{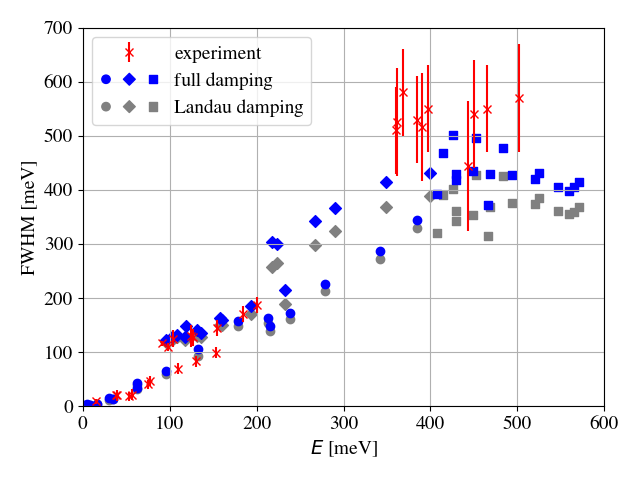}
	\caption{Magnon damping along $\bar{\Gamma}$-$\bar{X}$ and
          $\bar{\Gamma}$-$\bar{M}$ due to Landau (ordered, LSDA+$U$) and non-Landau
          mechanisms (disordered, LSDA+$U$). The Landau damping (grey markers) and the full
          damping (including Landau, disorder and multi-magnon
          processes) depicted with blue markers are compared to
          experimental data \cite{Zakeri2021}. Different magnon modes are represented by
          different marker shapes.} 
	\label{fig_damp}
\end{figure}

In order to quantify these contributions we write the Heisenberg Hamiltonian as follows:
\begin{align}\label{eqn_H}
	H = \sum_{\vb*{k}}\sum_{ij}a_i^\dagger(\vb*{k})T_{ij}(\vb*{k})a_j(\vb*{k}) + H_4 + H_{SOC}
\end{align}
Here, the first term describes the non-interacting bosons with the torque matrix $T$ while $H_4$ is the first term beyond the linearization of the Holstein-Primakoff transformation describing the four magnon processes involving the interaction with the magnon bath, cf. Fig. \ref{fig_damping_channels}~c. $H_\text{SOC}$ represents the three-magnon processes enabled by the SOC, cf. Fig. \ref{fig_damping_channels}~d. For the latter Hamiltonian, the exemplary  Dzyaloshinskii-Moriya interaction (DMI) form is utilized. Both $H_4$ as well as $H_{\text{SOC}}$ are treated as perturbations and the corresponding magnon decay rates can be calculated using Fermi's golden rule
$
	\Gamma_{sc}^{i\rightarrow f}  = 2\pi \rho(E_f) \abs{\matrixel{f_{sc}}{H_{sc}}{i_{sc}}}^2 
$
with $sc\in\{4,\text{SOC}\}$. 
Here, both the initial and final state for the four-magnon process ($\ket{i_4}$ and $\ket{f_4}$) are two-magnon states. For the three-magnon process, the initial state $\ket{i_\text{SOC}} = \mathcal{A}^\dagger\ket{0}$ is a single magnon state ($\mathcal{A}^\dagger$ being magnon creation operator acting on the vacuum state) and the final state $\ket{f_\text{SOC}}$ is a two-magnon state. A detailed exposition of the formalism and the results is given in the supplementary note I \cite{supplement}. The main observation is that the magnon-magnon induced decay rate is orders of magnitude smaller than the Landau damping: $\Gamma_{4} < 5\unit{meV}$ and $\Gamma_{SOC} < 1\unit{meV}$.\\
%To obtain the damping of one specific magnon mode, we sum over all possible final states that satisfy the momentum and energy conservation. % conservation exactly and the energy conservation up to a small energy tolerance $\epsilon$.
%For the four-magnon processes an additional summation over the scattering partner is needed. As there is no full \emph{ab initio} treatment of the SPHREELS process so far, we have to model the occupation probability for the magnon bath created in the SPHREELS process. We assume that all possible magnon modes are generated at an equal constant rate in the experiment but account for the higher decay rates of more energetic modes. The order of magnitude of the resulting magnon damping is independent of the details occupation function. A detailed exposition of the formalism and the results is given in the supplementary material. The major observation is that the magnon-magnon induced decay rate is orders of magnitude smaller than the Landau damping: $\Gamma_{4} < 5$meV and $\Gamma_{SOC} < 1$meV.
Lastly, we show this is not the case with the disorder induced damping \cite{Paischer2021a}. We find that the disorder induced damping is sensitive to the shape of the magnon mode and can differently affect two modes of similar energy and momentum, leading to a \enquote{mode selective damping} and constituting an attractive linewidth engineering approach, cf. supplementary note II in \cite{supplement}. %A spin-wave is associated with a Bloch wave delocalized over the entire lattice. Random impurity atoms break the translational symmetry of the system and the Bloch wave can decay into waves of other momenta with the same energy. In our recent work  \cite{Buczek2016,Paischer2021a,Paischer2021}, we have generalized the CPA formalism to systematically account for the non-diagonal disorder of the Heisenberg Hamiltonian in low dimensional systems featuring complex unit cells in order to describe the spin-wave propagation and scattering in disordered nanostructures. In general, the magnons below 200meV are weakly affected by the disorder and the attenuation increases with the magnon energy. However, we find that the disorder induced damping is sensitive to the shape of the magnon mode and can differently affect two modes of similar energy and momentum, leading to a \enquote{mode selective damping} and constituting an attractive linewidth engineering approach, cf. supplementary material.
The total magnon damping for Co/Ir compared with the experimental data
is shown in Fig. \ref{fig_damp}.   Despite the large error bars of the highest energy mangon mode, there is a clear trend showing that the Landau damping itself cannot account for the width of the peaks. Their linewidth is clearly underestimated and the inclusion of the non-Landau damping channels, dominated by the disorder induced scattering, accounts for a substantial part of the missing line
width. However, the damping is still underestimated which we belive might
result from the coupling of the transverse magnons to the longitudinal
spin dynamics \cite{Buczek2020} arising due to the SOC but not
included in this work. \\
We find the same magnitude of non-Landau damping in Co/Cu. Hence, we find that the magnon damping in Co/Cu is generally overestimated by up to 50meV. The reason for this might be the missing damping of electronic states within the LSDA+$U$ as mentioned before.
%While the agreement of theory and experiment for magnons with low energy is not changed, the higher energy magnons suffer visibly from non-Landau damping mechanisms, primarily disorder induced scattering. While for the latter, the gap between the experimental and theoretical values is narrowed, there is still a sizable disagreement. There are a few possible reasons for this. First, one can see that the agreement between electronic spectra obtained from LSDA+$U$ and the LSDA+electron-magnon scattering differ in Co/Ir much more than in the case of Co/Cu. This might be a reason for the fact that Co/Cu is well described while this is not quite the case for Co/Ir. Second, there might yet be other damping mechanisms at play in Co/Ir, most likely due to the higher spin-orbit coupling caused by the Ir substrate. The SOC enters in our approach only through the DMI parameters used for the multi-magnon processes whose impact is minor, as discussed. However, the SOC allows for another damping channel not considered so far, which are electronic excitations between states with equal spin projection.
\newline In summary, using a first principles approach, we uncovered
an intricate picture of spin dynamics and its damping in itinerant 3d
magnetic nanostructures governed by a fine interplay between
non-trivial correlation effects dominated by electron-magnon
scattering and disorder. We believe that the electron-magnon interaction might be a generic effect and an indispensable ingredient in the description of all itinerant magnets. Furthermore, we established a quantitative
hierarchy of attenuation mechanisms dominated by the Landau channel
and disorder with the multi-magnon processes playing a secondary
role. 

S.P. is recipient of a DOC Fellowship of the Austrian Academy of
Sciences at the Institute of mathematics, physics, space research and
materials sciences. I.M. and P.B. gratefully acknowledge financial
support from the DFG-LAV grant \enquote{SPINELS} and HSP grant
\enquote{DEUM}. A.E. acknowledges funding by Fonds zur Förderung der
Wissenschaftlichen Forschung (FWF) Grant No. I 5384. The research of Kh.Z. has financially been supported by DFG through grants ZA 902/7-1 and ZA 902/8-1. We thank A. Marmodoro for interesting discussions and comments.
\bibliographystyle{apsrev4-2}
%\end{thebibliography}
\bibliography{gw,magnetism, Quellen}

\begin{thebibliography}{45}%
\makeatletter
\providecommand \@ifxundefined [1]{%
 \@ifx{#1\undefined}
}%
\providecommand \@ifnum [1]{%
 \ifnum #1\expandafter \@firstoftwo
 \else \expandafter \@secondoftwo
 \fi
}%
\providecommand \@ifx [1]{%
 \ifx #1\expandafter \@firstoftwo
 \else \expandafter \@secondoftwo
 \fi
}%
\providecommand \natexlab [1]{#1}%
\providecommand \enquote  [1]{``#1''}%
\providecommand \bibnamefont  [1]{#1}%
\providecommand \bibfnamefont [1]{#1}%
\providecommand \citenamefont [1]{#1}%
\providecommand \href@noop [0]{\@secondoftwo}%
\providecommand \href [0]{\begingroup \@sanitize@url \@href}%
\providecommand \@href[1]{\@@startlink{#1}\@@href}%
\providecommand \@@href[1]{\endgroup#1\@@endlink}%
\providecommand \@sanitize@url [0]{\catcode `\\12\catcode `\$12\catcode
  `\&12\catcode `\#12\catcode `\^12\catcode `\_12\catcode `\%12\relax}%
\providecommand \@@startlink[1]{}%
\providecommand \@@endlink[0]{}%
\providecommand \url  [0]{\begingroup\@sanitize@url \@url }%
\providecommand \@url [1]{\endgroup\@href {#1}{\urlprefix }}%
\providecommand \urlprefix  [0]{URL }%
\providecommand \Eprint [0]{\href }%
\providecommand \doibase [0]{https://doi.org/}%
\providecommand \selectlanguage [0]{\@gobble}%
\providecommand \bibinfo  [0]{\@secondoftwo}%
\providecommand \bibfield  [0]{\@secondoftwo}%
\providecommand \translation [1]{[#1]}%
\providecommand \BibitemOpen [0]{}%
\providecommand \bibitemStop [0]{}%
\providecommand \bibitemNoStop [0]{.\EOS\space}%
\providecommand \EOS [0]{\spacefactor3000\relax}%
\providecommand \BibitemShut  [1]{\csname bibitem#1\endcsname}%
\let\auto@bib@innerbib\@empty
%</preamble>
\bibitem [{\citenamefont {Yang}\ \emph {et~al.}(2021)\citenamefont {Yang},
  \citenamefont {Liu}, \citenamefont {Bi},\ and\ \citenamefont
  {Deng}}]{Yang2021}%
  \BibitemOpen
  \bibfield  {author} {\bibinfo {author} {\bibfnamefont {Y.}~\bibnamefont
  {Yang}}, \bibinfo {author} {\bibfnamefont {T.}~\bibnamefont {Liu}}, \bibinfo
  {author} {\bibfnamefont {L.}~\bibnamefont {Bi}},\ and\ \bibinfo {author}
  {\bibfnamefont {L.}~\bibnamefont {Deng}},\ }\href
  {https://doi.org/10.1016/j.jallcom.2020.158235} {\bibfield  {journal}
  {\bibinfo  {journal} {Journal of Alloys and Compounds}\ }\textbf {\bibinfo
  {volume} {860}},\ \bibinfo {pages} {158235} (\bibinfo {year}
  {2021})}\BibitemShut {NoStop}%
\bibitem [{\citenamefont {Natterer}\ \emph {et~al.}(2017)\citenamefont
  {Natterer}, \citenamefont {Yang}, \citenamefont {Paul}, \citenamefont
  {Willke}, \citenamefont {Choi}, \citenamefont {Greber}, \citenamefont
  {Heinrich},\ and\ \citenamefont {Lutz}}]{Natterer2017}%
  \BibitemOpen
  \bibfield  {author} {\bibinfo {author} {\bibfnamefont {F.~D.}\ \bibnamefont
  {Natterer}}, \bibinfo {author} {\bibfnamefont {K.}~\bibnamefont {Yang}},
  \bibinfo {author} {\bibfnamefont {W.}~\bibnamefont {Paul}}, \bibinfo {author}
  {\bibfnamefont {P.}~\bibnamefont {Willke}}, \bibinfo {author} {\bibfnamefont
  {T.}~\bibnamefont {Choi}}, \bibinfo {author} {\bibfnamefont {T.}~\bibnamefont
  {Greber}}, \bibinfo {author} {\bibfnamefont {A.~J.}\ \bibnamefont
  {Heinrich}},\ and\ \bibinfo {author} {\bibfnamefont {C.~P.}\ \bibnamefont
  {Lutz}},\ }\href {https://doi.org/10.1038/nature21371} {\bibfield  {journal}
  {\bibinfo  {journal} {Nature}\ }\textbf {\bibinfo {volume} {543}},\ \bibinfo
  {pages} {226} (\bibinfo {year} {2017})}\BibitemShut {NoStop}%
\bibitem [{\citenamefont {Guo}\ \emph {et~al.}(2019)\citenamefont {Guo},
  \citenamefont {Gu}, \citenamefont {Zhu},\ and\ \citenamefont
  {Sun}}]{Guo2019}%
  \BibitemOpen
  \bibfield  {author} {\bibinfo {author} {\bibfnamefont {L.}~\bibnamefont
  {Guo}}, \bibinfo {author} {\bibfnamefont {X.}~\bibnamefont {Gu}}, \bibinfo
  {author} {\bibfnamefont {X.}~\bibnamefont {Zhu}},\ and\ \bibinfo {author}
  {\bibfnamefont {X.}~\bibnamefont {Sun}},\ }\href
  {https://doi.org/10.1002/adma.201805355} {\bibfield  {journal} {\bibinfo
  {journal} {Advanced Materials}\ }\textbf {\bibinfo {volume} {31}},\ \bibinfo
  {pages} {1805355} (\bibinfo {year} {2019})}\BibitemShut {NoStop}%
\bibitem [{\citenamefont {Jia}\ \emph {et~al.}(2021)\citenamefont {Jia},
  \citenamefont {Chen}, \citenamefont {Schäffer},\ and\ \citenamefont
  {Berakdar}}]{Jia2021}%
  \BibitemOpen
  \bibfield  {author} {\bibinfo {author} {\bibfnamefont {C.}~\bibnamefont
  {Jia}}, \bibinfo {author} {\bibfnamefont {M.}~\bibnamefont {Chen}}, \bibinfo
  {author} {\bibfnamefont {A.~F.}\ \bibnamefont {Schäffer}},\ and\ \bibinfo
  {author} {\bibfnamefont {J.}~\bibnamefont {Berakdar}},\ }\bibfield  {journal}
  {\bibinfo  {journal} {npj Computational Materials}\ }\textbf {\bibinfo
  {volume} {7}},\ \href {https://doi.org/10.1038/s41524-021-00570-0}
  {10.1038/s41524-021-00570-0} (\bibinfo {year} {2021})\BibitemShut {NoStop}%
\bibitem [{\citenamefont {Hadri}\ \emph {et~al.}(2016)\citenamefont {Hadri},
  \citenamefont {Pirro}, \citenamefont {Lambert}, \citenamefont
  {Petit-Watelot}, \citenamefont {Quessab}, \citenamefont {Hehn}, \citenamefont
  {Montaigne}, \citenamefont {Malinowski},\ and\ \citenamefont
  {Mangin}}]{Hadri2016}%
  \BibitemOpen
  \bibfield  {author} {\bibinfo {author} {\bibfnamefont {M.~S.~E.}\
  \bibnamefont {Hadri}}, \bibinfo {author} {\bibfnamefont {P.}~\bibnamefont
  {Pirro}}, \bibinfo {author} {\bibfnamefont {C.-H.}\ \bibnamefont {Lambert}},
  \bibinfo {author} {\bibfnamefont {S.}~\bibnamefont {Petit-Watelot}}, \bibinfo
  {author} {\bibfnamefont {Y.}~\bibnamefont {Quessab}}, \bibinfo {author}
  {\bibfnamefont {M.}~\bibnamefont {Hehn}}, \bibinfo {author} {\bibfnamefont
  {F.}~\bibnamefont {Montaigne}}, \bibinfo {author} {\bibfnamefont
  {G.}~\bibnamefont {Malinowski}},\ and\ \bibinfo {author} {\bibfnamefont
  {S.}~\bibnamefont {Mangin}},\ }\bibfield  {journal} {\bibinfo  {journal}
  {Physical Review B}\ }\textbf {\bibinfo {volume} {94}},\ \href
  {https://doi.org/10.1103/physrevb.94.064412} {10.1103/physrevb.94.064412}
  (\bibinfo {year} {2016})\BibitemShut {NoStop}%
\bibitem [{\citenamefont {Sobota}\ \emph {et~al.}(2021)\citenamefont {Sobota},
  \citenamefont {He},\ and\ \citenamefont {Shen}}]{Sobota2021}%
  \BibitemOpen
  \bibfield  {author} {\bibinfo {author} {\bibfnamefont {J.~A.}\ \bibnamefont
  {Sobota}}, \bibinfo {author} {\bibfnamefont {Y.}~\bibnamefont {He}},\ and\
  \bibinfo {author} {\bibfnamefont {Z.-X.}\ \bibnamefont {Shen}},\ }\bibfield
  {journal} {\bibinfo  {journal} {Reviews of Modern Physics}\ }\textbf
  {\bibinfo {volume} {93}},\ \href
  {https://doi.org/10.1103/revmodphys.93.025006} {10.1103/revmodphys.93.025006}
  (\bibinfo {year} {2021})\BibitemShut {NoStop}%
\bibitem [{\citenamefont {Qin}\ \emph {et~al.}(2019)\citenamefont {Qin},
  \citenamefont {Tsurkan}, \citenamefont {Ernst},\ and\ \citenamefont
  {Zakeri}}]{Qin2019}%
  \BibitemOpen
  \bibfield  {author} {\bibinfo {author} {\bibfnamefont {H.~J.}\ \bibnamefont
  {Qin}}, \bibinfo {author} {\bibfnamefont {S.}~\bibnamefont {Tsurkan}},
  \bibinfo {author} {\bibfnamefont {A.}~\bibnamefont {Ernst}},\ and\ \bibinfo
  {author} {\bibfnamefont {K.}~\bibnamefont {Zakeri}},\ }\href
  {https://doi.org/10.1103/physrevlett.123.257202} {\bibfield  {journal}
  {\bibinfo  {journal} {Physical Review Letters}\ }\textbf {\bibinfo {volume}
  {123}},\ \bibinfo {pages} {257202} (\bibinfo {year} {2019})}\BibitemShut
  {NoStop}%
\bibitem [{\citenamefont {Zakeri}\ \emph {et~al.}(2021)\citenamefont {Zakeri},
  \citenamefont {Hjelt}, \citenamefont {Maznichenko}, \citenamefont {Buczek},\
  and\ \citenamefont {Ernst}}]{Zakeri2021}%
  \BibitemOpen
  \bibfield  {author} {\bibinfo {author} {\bibfnamefont {K.}~\bibnamefont
  {Zakeri}}, \bibinfo {author} {\bibfnamefont {A.}~\bibnamefont {Hjelt}},
  \bibinfo {author} {\bibfnamefont {I.~V.}\ \bibnamefont {Maznichenko}},
  \bibinfo {author} {\bibfnamefont {P.}~\bibnamefont {Buczek}},\ and\ \bibinfo
  {author} {\bibfnamefont {A.}~\bibnamefont {Ernst}},\ }\href
  {https://doi.org/10.1103/PhysRevLett.126.177203} {\bibfield  {journal}
  {\bibinfo  {journal} {Phys. Rev. Lett.}\ }\textbf {\bibinfo {volume} {126}},\
  \bibinfo {pages} {177203} (\bibinfo {year} {2021})}\BibitemShut {NoStop}%
\bibitem [{\citenamefont {Essenberger}\ \emph {et~al.}(2016)\citenamefont
  {Essenberger}, \citenamefont {Sanna}, \citenamefont {Buczek}, \citenamefont
  {Ernst}, \citenamefont {Sandratskii},\ and\ \citenamefont
  {Gross}}]{Essenberger2016}%
  \BibitemOpen
  \bibfield  {author} {\bibinfo {author} {\bibfnamefont {F.}~\bibnamefont
  {Essenberger}}, \bibinfo {author} {\bibfnamefont {A.}~\bibnamefont {Sanna}},
  \bibinfo {author} {\bibfnamefont {P.}~\bibnamefont {Buczek}}, \bibinfo
  {author} {\bibfnamefont {A.}~\bibnamefont {Ernst}}, \bibinfo {author}
  {\bibfnamefont {L.}~\bibnamefont {Sandratskii}},\ and\ \bibinfo {author}
  {\bibfnamefont {E.~K.~U.}\ \bibnamefont {Gross}},\ }\href
  {https://link.aps.org/doi/10.1103/PhysRevB.94.014503} {\bibfield  {journal}
  {\bibinfo  {journal} {Phys. Rev. B}\ }\textbf {\bibinfo {volume} {94}},\
  \bibinfo {pages} {014503} (\bibinfo {year} {2016})}\BibitemShut {NoStop}%
\bibitem [{\citenamefont {Tusche}\ \emph {et~al.}(2018)\citenamefont {Tusche},
  \citenamefont {Ellguth}, \citenamefont {Feyer}, \citenamefont {Krasyuk},
  \citenamefont {Wiemann}, \citenamefont {Henk}, \citenamefont {Schneider},\
  and\ \citenamefont {Kirschner}}]{Tusche2018}%
  \BibitemOpen
  \bibfield  {author} {\bibinfo {author} {\bibfnamefont {C.}~\bibnamefont
  {Tusche}}, \bibinfo {author} {\bibfnamefont {M.}~\bibnamefont {Ellguth}},
  \bibinfo {author} {\bibfnamefont {V.}~\bibnamefont {Feyer}}, \bibinfo
  {author} {\bibfnamefont {A.}~\bibnamefont {Krasyuk}}, \bibinfo {author}
  {\bibfnamefont {C.}~\bibnamefont {Wiemann}}, \bibinfo {author} {\bibfnamefont
  {J.}~\bibnamefont {Henk}}, \bibinfo {author} {\bibfnamefont {C.~M.}\
  \bibnamefont {Schneider}},\ and\ \bibinfo {author} {\bibfnamefont
  {J.}~\bibnamefont {Kirschner}},\ }\bibfield  {journal} {\bibinfo  {journal}
  {Nature Communications}\ }\textbf {\bibinfo {volume} {9}},\ \href
  {https://doi.org/10.1038/s41467-018-05960-5} {10.1038/s41467-018-05960-5}
  (\bibinfo {year} {2018})\BibitemShut {NoStop}%
\bibitem [{\citenamefont {Schmidt}\ \emph {et~al.}(2010)\citenamefont
  {Schmidt}, \citenamefont {Pickel}, \citenamefont {Donath}, \citenamefont
  {Buczek}, \citenamefont {Ernst}, \citenamefont {Zhukov}, \citenamefont
  {Echenique}, \citenamefont {Sandratskii}, \citenamefont {Chulkov},\ and\
  \citenamefont {Weinelt}}]{Schmidt2010a}%
  \BibitemOpen
  \bibfield  {author} {\bibinfo {author} {\bibfnamefont {A.~B.}\ \bibnamefont
  {Schmidt}}, \bibinfo {author} {\bibfnamefont {M.}~\bibnamefont {Pickel}},
  \bibinfo {author} {\bibfnamefont {M.}~\bibnamefont {Donath}}, \bibinfo
  {author} {\bibfnamefont {P.}~\bibnamefont {Buczek}}, \bibinfo {author}
  {\bibfnamefont {A.}~\bibnamefont {Ernst}}, \bibinfo {author} {\bibfnamefont
  {V.~P.}\ \bibnamefont {Zhukov}}, \bibinfo {author} {\bibfnamefont {P.~M.}\
  \bibnamefont {Echenique}}, \bibinfo {author} {\bibfnamefont {L.~M.}\
  \bibnamefont {Sandratskii}}, \bibinfo {author} {\bibfnamefont {E.~V.}\
  \bibnamefont {Chulkov}},\ and\ \bibinfo {author} {\bibfnamefont
  {M.}~\bibnamefont {Weinelt}},\ }\href
  {http://link.aps.org/doi/10.1103/PhysRevLett.105.197401} {\bibfield
  {journal} {\bibinfo  {journal} {Phys. Rev. Lett.}\ }\textbf {\bibinfo
  {volume} {105}},\ \bibinfo {pages} {197401} (\bibinfo {year}
  {2010})}\BibitemShut {NoStop}%
\bibitem [{\citenamefont {Chumak}\ \emph {et~al.}(2017)\citenamefont {Chumak},
  \citenamefont {Serga},\ and\ \citenamefont {Hillebrands}}]{Chumak2017}%
  \BibitemOpen
  \bibfield  {author} {\bibinfo {author} {\bibfnamefont {A.~V.}\ \bibnamefont
  {Chumak}}, \bibinfo {author} {\bibfnamefont {A.~A.}\ \bibnamefont {Serga}},\
  and\ \bibinfo {author} {\bibfnamefont {B.}~\bibnamefont {Hillebrands}},\
  }\href {https://doi.org/10.1088/1361-6463/aa6a65} {\bibfield  {journal}
  {\bibinfo  {journal} {Journal of Physics D: Applied Physics}\ }\textbf
  {\bibinfo {volume} {50}},\ \bibinfo {pages} {244001} (\bibinfo {year}
  {2017})}\BibitemShut {NoStop}%
\bibitem [{\citenamefont {Varykhalov}\ \emph {et~al.}(2005)\citenamefont
  {Varykhalov}, \citenamefont {Shikin}, \citenamefont {Gudat}, \citenamefont
  {Moras}, \citenamefont {Grazioli}, \citenamefont {Carbone},\ and\
  \citenamefont {Rader}}]{Varykhalov2005}%
  \BibitemOpen
  \bibfield  {author} {\bibinfo {author} {\bibfnamefont {A.}~\bibnamefont
  {Varykhalov}}, \bibinfo {author} {\bibfnamefont {A.~M.}\ \bibnamefont
  {Shikin}}, \bibinfo {author} {\bibfnamefont {W.}~\bibnamefont {Gudat}},
  \bibinfo {author} {\bibfnamefont {P.}~\bibnamefont {Moras}}, \bibinfo
  {author} {\bibfnamefont {C.}~\bibnamefont {Grazioli}}, \bibinfo {author}
  {\bibfnamefont {C.}~\bibnamefont {Carbone}},\ and\ \bibinfo {author}
  {\bibfnamefont {O.}~\bibnamefont {Rader}},\ }\bibfield  {journal} {\bibinfo
  {journal} {Physical Review Letters}\ }\textbf {\bibinfo {volume} {95}},\
  \href {https://doi.org/10.1103/physrevlett.95.247601}
  {10.1103/physrevlett.95.247601} (\bibinfo {year} {2005})\BibitemShut
  {NoStop}%
\bibitem [{\citenamefont {Kotliar}\ \emph {et~al.}(2006)\citenamefont
  {Kotliar}, \citenamefont {Savrasov}, \citenamefont {Haule}, \citenamefont
  {Oudovenko}, \citenamefont {Parcollet},\ and\ \citenamefont
  {Marianetti}}]{Kotliar2006}%
  \BibitemOpen
  \bibfield  {author} {\bibinfo {author} {\bibfnamefont {G.}~\bibnamefont
  {Kotliar}}, \bibinfo {author} {\bibfnamefont {S.~Y.}\ \bibnamefont
  {Savrasov}}, \bibinfo {author} {\bibfnamefont {K.}~\bibnamefont {Haule}},
  \bibinfo {author} {\bibfnamefont {V.~S.}\ \bibnamefont {Oudovenko}}, \bibinfo
  {author} {\bibfnamefont {O.}~\bibnamefont {Parcollet}},\ and\ \bibinfo
  {author} {\bibfnamefont {C.~A.}\ \bibnamefont {Marianetti}},\ }\href
  {http://link.aps.org/abstract/RMP/v78/p865} {\bibfield  {journal} {\bibinfo
  {journal} {Rev. Mod. Phys.}\ }\textbf {\bibinfo {volume} {78}},\ \bibinfo
  {pages} {865} (\bibinfo {year} {2006})}\BibitemShut {NoStop}%
\bibitem [{\citenamefont {Monastra}\ \emph {et~al.}(2002)\citenamefont
  {Monastra}, \citenamefont {Manghi}, \citenamefont {Rozzi}, \citenamefont
  {Arcangeli}, \citenamefont {Wetli}, \citenamefont {Neff}, \citenamefont
  {Greber},\ and\ \citenamefont {Osterwalder}}]{Monastra2002}%
  \BibitemOpen
  \bibfield  {author} {\bibinfo {author} {\bibfnamefont {S.}~\bibnamefont
  {Monastra}}, \bibinfo {author} {\bibfnamefont {F.}~\bibnamefont {Manghi}},
  \bibinfo {author} {\bibfnamefont {C.~A.}\ \bibnamefont {Rozzi}}, \bibinfo
  {author} {\bibfnamefont {C.}~\bibnamefont {Arcangeli}}, \bibinfo {author}
  {\bibfnamefont {E.}~\bibnamefont {Wetli}}, \bibinfo {author} {\bibfnamefont
  {H.-J.}\ \bibnamefont {Neff}}, \bibinfo {author} {\bibfnamefont
  {T.}~\bibnamefont {Greber}},\ and\ \bibinfo {author} {\bibfnamefont
  {J.}~\bibnamefont {Osterwalder}},\ }\href
  {https://doi.org/10.1103/physrevlett.88.236402} {\bibfield  {journal}
  {\bibinfo  {journal} {Physical Review Letters}\ }\textbf {\bibinfo {volume}
  {88}},\ \bibinfo {pages} {236402} (\bibinfo {year} {2002})}\BibitemShut
  {NoStop}%
\bibitem [{\citenamefont {{Costa, Jr.}}\ \emph {et~al.}(2004)\citenamefont
  {{Costa, Jr.}}, \citenamefont {Muniz},\ and\ \citenamefont
  {Mills}}]{Costa2004}%
  \BibitemOpen
  \bibfield  {author} {\bibinfo {author} {\bibfnamefont {A.~T.}\ \bibnamefont
  {{Costa, Jr.}}}, \bibinfo {author} {\bibfnamefont {R.~B.}\ \bibnamefont
  {Muniz}},\ and\ \bibinfo {author} {\bibfnamefont {D.~L.}\ \bibnamefont
  {Mills}},\ }\href {http://link.aps.org/abstract/PRB/v69/e064413} {\bibfield
  {journal} {\bibinfo  {journal} {Physical Review B (Condensed Matter and
  Materials Physics)}\ }\textbf {\bibinfo {volume} {69}},\ \bibinfo {eid}
  {064413} (\bibinfo {year} {2004})}\BibitemShut {NoStop}%
\bibitem [{\citenamefont {Taroni}\ \emph {et~al.}(2011)\citenamefont {Taroni},
  \citenamefont {Bergman}, \citenamefont {Bergqvist}, \citenamefont
  {Hellsvik},\ and\ \citenamefont {Eriksson}}]{Taroni2011}%
  \BibitemOpen
  \bibfield  {author} {\bibinfo {author} {\bibfnamefont {A.}~\bibnamefont
  {Taroni}}, \bibinfo {author} {\bibfnamefont {A.}~\bibnamefont {Bergman}},
  \bibinfo {author} {\bibfnamefont {L.}~\bibnamefont {Bergqvist}}, \bibinfo
  {author} {\bibfnamefont {J.}~\bibnamefont {Hellsvik}},\ and\ \bibinfo
  {author} {\bibfnamefont {O.}~\bibnamefont {Eriksson}},\ }\bibfield  {journal}
  {\bibinfo  {journal} {Physical Review Letters}\ }\textbf {\bibinfo {volume}
  {107}},\ \href {https://doi.org/10.1103/physrevlett.107.037202}
  {10.1103/physrevlett.107.037202} (\bibinfo {year} {2011})\BibitemShut
  {NoStop}%
\bibitem [{\citenamefont {Chen}\ \emph {et~al.}(2017)\citenamefont {Chen},
  \citenamefont {Zakeri}, \citenamefont {Ernst}, \citenamefont {Qin},
  \citenamefont {Meng},\ and\ \citenamefont {Kirschner}}]{Chen2017}%
  \BibitemOpen
  \bibfield  {author} {\bibinfo {author} {\bibfnamefont {Y.-J.}\ \bibnamefont
  {Chen}}, \bibinfo {author} {\bibfnamefont {K.}~\bibnamefont {Zakeri}},
  \bibinfo {author} {\bibfnamefont {A.}~\bibnamefont {Ernst}}, \bibinfo
  {author} {\bibfnamefont {H.~J.}\ \bibnamefont {Qin}}, \bibinfo {author}
  {\bibfnamefont {Y.}~\bibnamefont {Meng}},\ and\ \bibinfo {author}
  {\bibfnamefont {J.}~\bibnamefont {Kirschner}},\ }\href
  {https://doi.org/10.1103/physrevlett.119.267201} {\bibfield  {journal}
  {\bibinfo  {journal} {Physical Review Letters}\ }\textbf {\bibinfo {volume}
  {119}},\ \bibinfo {pages} {267201} (\bibinfo {year} {2017})}\BibitemShut
  {NoStop}%
\bibitem [{\citenamefont {Janas}\ \emph {et~al.}(2022)\citenamefont {Janas},
  \citenamefont {Droghetti}, \citenamefont {Ponzoni}, \citenamefont
  {Cojocariu}, \citenamefont {Jugovac}, \citenamefont {Feyer}, \citenamefont
  {Radonji{\'{c}}}, \citenamefont {Rungger}, \citenamefont {Chioncel},
  \citenamefont {Zamborlini},\ and\ \citenamefont {Cinchetti}}]{Janas2022}%
  \BibitemOpen
  \bibfield  {author} {\bibinfo {author} {\bibfnamefont {D.~M.}\ \bibnamefont
  {Janas}}, \bibinfo {author} {\bibfnamefont {A.}~\bibnamefont {Droghetti}},
  \bibinfo {author} {\bibfnamefont {S.}~\bibnamefont {Ponzoni}}, \bibinfo
  {author} {\bibfnamefont {I.}~\bibnamefont {Cojocariu}}, \bibinfo {author}
  {\bibfnamefont {M.}~\bibnamefont {Jugovac}}, \bibinfo {author} {\bibfnamefont
  {V.}~\bibnamefont {Feyer}}, \bibinfo {author} {\bibfnamefont {M.~M.}\
  \bibnamefont {Radonji{\'{c}}}}, \bibinfo {author} {\bibfnamefont
  {I.}~\bibnamefont {Rungger}}, \bibinfo {author} {\bibfnamefont
  {L.}~\bibnamefont {Chioncel}}, \bibinfo {author} {\bibfnamefont
  {G.}~\bibnamefont {Zamborlini}},\ and\ \bibinfo {author} {\bibfnamefont
  {M.}~\bibnamefont {Cinchetti}},\ }\href
  {https://doi.org/10.1002/adma.202205698} {\bibfield  {journal} {\bibinfo
  {journal} {Advanced Materials}\ }\textbf {\bibinfo {volume} {35}},\ \bibinfo
  {pages} {2205698} (\bibinfo {year} {2022})}\BibitemShut {NoStop}%
\bibitem [{\citenamefont {Paischer}\ \emph {et~al.}(2023)\citenamefont
  {Paischer}, \citenamefont {Vignale}, \citenamefont {Katsnelson},
  \citenamefont {Ernst},\ and\ \citenamefont {Buczek}}]{Paischer2023}%
  \BibitemOpen
  \bibfield  {author} {\bibinfo {author} {\bibfnamefont {S.}~\bibnamefont
  {Paischer}}, \bibinfo {author} {\bibfnamefont {G.}~\bibnamefont {Vignale}},
  \bibinfo {author} {\bibfnamefont {M.~I.}\ \bibnamefont {Katsnelson}},
  \bibinfo {author} {\bibfnamefont {A.}~\bibnamefont {Ernst}},\ and\ \bibinfo
  {author} {\bibfnamefont {P.~A.}\ \bibnamefont {Buczek}},\ }\href
  {https://doi.org/10.1103/physrevb.107.134410} {\bibfield  {journal} {\bibinfo
   {journal} {Physical Review B}\ }\textbf {\bibinfo {volume} {107}},\ \bibinfo
  {pages} {134410} (\bibinfo {year} {2023})}\BibitemShut {NoStop}%
\bibitem [{\citenamefont {Buczek}\ \emph {et~al.}(2011)\citenamefont {Buczek},
  \citenamefont {Ernst},\ and\ \citenamefont {Sandratskii}}]{Buczek2011a}%
  \BibitemOpen
  \bibfield  {author} {\bibinfo {author} {\bibfnamefont {P.}~\bibnamefont
  {Buczek}}, \bibinfo {author} {\bibfnamefont {A.}~\bibnamefont {Ernst}},\ and\
  \bibinfo {author} {\bibfnamefont {L.~M.}\ \bibnamefont {Sandratskii}},\
  }\href {http://link.aps.org/doi/10.1103/PhysRevB.84.174418} {\bibfield
  {journal} {\bibinfo  {journal} {Phys. Rev. B}\ }\textbf {\bibinfo {volume}
  {84}},\ \bibinfo {pages} {174418} (\bibinfo {year} {2011})}\BibitemShut
  {NoStop}%
\bibitem [{\citenamefont {Qin}\ \emph {et~al.}(2015)\citenamefont {Qin},
  \citenamefont {Zakeri}, \citenamefont {Ernst}, \citenamefont {Sandratskii},
  \citenamefont {Buczek}, \citenamefont {Marmodoro}, \citenamefont {Chuang},
  \citenamefont {Zhang},\ and\ \citenamefont {Kirschner}}]{Qin2015}%
  \BibitemOpen
  \bibfield  {author} {\bibinfo {author} {\bibfnamefont {H.~J.}\ \bibnamefont
  {Qin}}, \bibinfo {author} {\bibfnamefont {K.}~\bibnamefont {Zakeri}},
  \bibinfo {author} {\bibfnamefont {A.}~\bibnamefont {Ernst}}, \bibinfo
  {author} {\bibfnamefont {L.~M.}\ \bibnamefont {Sandratskii}}, \bibinfo
  {author} {\bibfnamefont {P.}~\bibnamefont {Buczek}}, \bibinfo {author}
  {\bibfnamefont {A.}~\bibnamefont {Marmodoro}}, \bibinfo {author}
  {\bibfnamefont {T.~H.}\ \bibnamefont {Chuang}}, \bibinfo {author}
  {\bibfnamefont {Y.}~\bibnamefont {Zhang}},\ and\ \bibinfo {author}
  {\bibfnamefont {J.}~\bibnamefont {Kirschner}},\ }\href
  {http://dx.doi.org/10.1038/ncomms7126} {\bibfield  {journal} {\bibinfo
  {journal} {Nat Commun}\ }\textbf {\bibinfo {volume} {6}},\ \bibinfo {pages}
  {6126} (\bibinfo {year} {2015})}\BibitemShut {NoStop}%
\bibitem [{\citenamefont {Kaganov}\ and\ \citenamefont
  {Chubukov}(1987)}]{Kaganov1987}%
  \BibitemOpen
  \bibfield  {author} {\bibinfo {author} {\bibfnamefont {M.~I.}\ \bibnamefont
  {Kaganov}}\ and\ \bibinfo {author} {\bibfnamefont {A.~V.}\ \bibnamefont
  {Chubukov}},\ }\href {https://doi.org/10.1070/pu1987v030n12abeh003065}
  {\bibfield  {journal} {\bibinfo  {journal} {Soviet Physics Uspekhi}\ }\textbf
  {\bibinfo {volume} {30}},\ \bibinfo {pages} {1015} (\bibinfo {year}
  {1987})}\BibitemShut {NoStop}%
\bibitem [{\citenamefont {Bergman}\ \emph {et~al.}(2010)\citenamefont
  {Bergman}, \citenamefont {Taroni}, \citenamefont {Bergqvist}, \citenamefont
  {Hellsvik}, \citenamefont {Hj\"orvarsson},\ and\ \citenamefont
  {Eriksson}}]{Bergman2010}%
  \BibitemOpen
  \bibfield  {author} {\bibinfo {author} {\bibfnamefont {A.}~\bibnamefont
  {Bergman}}, \bibinfo {author} {\bibfnamefont {A.}~\bibnamefont {Taroni}},
  \bibinfo {author} {\bibfnamefont {L.}~\bibnamefont {Bergqvist}}, \bibinfo
  {author} {\bibfnamefont {J.}~\bibnamefont {Hellsvik}}, \bibinfo {author}
  {\bibfnamefont {B.}~\bibnamefont {Hj\"orvarsson}},\ and\ \bibinfo {author}
  {\bibfnamefont {O.}~\bibnamefont {Eriksson}},\ }\bibfield  {journal}
  {\bibinfo  {journal} {Physical Review B}\ }\textbf {\bibinfo {volume} {81}},\
  \href {https://doi.org/10.1103/physrevb.81.144416}
  {10.1103/physrevb.81.144416} (\bibinfo {year} {2010})\BibitemShut {NoStop}%
\bibitem [{\citenamefont {Bergqvist}\ \emph {et~al.}(2013)\citenamefont
  {Bergqvist}, \citenamefont {Taroni}, \citenamefont {Bergman}, \citenamefont
  {Etz},\ and\ \citenamefont {Eriksson}}]{Bergqvist2013}%
  \BibitemOpen
  \bibfield  {author} {\bibinfo {author} {\bibfnamefont {L.}~\bibnamefont
  {Bergqvist}}, \bibinfo {author} {\bibfnamefont {A.}~\bibnamefont {Taroni}},
  \bibinfo {author} {\bibfnamefont {A.}~\bibnamefont {Bergman}}, \bibinfo
  {author} {\bibfnamefont {C.}~\bibnamefont {Etz}},\ and\ \bibinfo {author}
  {\bibfnamefont {O.}~\bibnamefont {Eriksson}},\ }\bibfield  {journal}
  {\bibinfo  {journal} {Physical Review B}\ }\textbf {\bibinfo {volume} {87}},\
  \href {https://doi.org/10.1103/physrevb.87.144401}
  {10.1103/physrevb.87.144401} (\bibinfo {year} {2013})\BibitemShut {NoStop}%
\bibitem [{\citenamefont {Paischer}\ \emph {et~al.}(2021)\citenamefont
  {Paischer}, \citenamefont {Buczek}, \citenamefont {Buczek}, \citenamefont
  {Eilmsteiner},\ and\ \citenamefont {Ernst}}]{Paischer2021a}%
  \BibitemOpen
  \bibfield  {author} {\bibinfo {author} {\bibfnamefont {S.}~\bibnamefont
  {Paischer}}, \bibinfo {author} {\bibfnamefont {P.~A.}\ \bibnamefont
  {Buczek}}, \bibinfo {author} {\bibfnamefont {N.}~\bibnamefont {Buczek}},
  \bibinfo {author} {\bibfnamefont {D.}~\bibnamefont {Eilmsteiner}},\ and\
  \bibinfo {author} {\bibfnamefont {A.}~\bibnamefont {Ernst}},\ }\href
  {https://doi.org/10.1103/PhysRevB.104.024403} {\bibfield  {journal} {\bibinfo
   {journal} {Phys. Rev. B}\ }\textbf {\bibinfo {volume} {104}},\ \bibinfo
  {pages} {024403} (\bibinfo {year} {2021})}\BibitemShut {NoStop}%
\bibitem [{\citenamefont {Hoffmann}\ \emph {et~al.}(2020)\citenamefont
  {Hoffmann}, \citenamefont {Ernst}, \citenamefont {Hergert}, \citenamefont
  {Antonov}, \citenamefont {Adeagbo}, \citenamefont {Geilhufe},\ and\
  \citenamefont {Hamed}}]{Hoffmann2020}%
  \BibitemOpen
  \bibfield  {author} {\bibinfo {author} {\bibfnamefont {M.}~\bibnamefont
  {Hoffmann}}, \bibinfo {author} {\bibfnamefont {A.}~\bibnamefont {Ernst}},
  \bibinfo {author} {\bibfnamefont {W.}~\bibnamefont {Hergert}}, \bibinfo
  {author} {\bibfnamefont {V.~N.}\ \bibnamefont {Antonov}}, \bibinfo {author}
  {\bibfnamefont {W.~A.}\ \bibnamefont {Adeagbo}}, \bibinfo {author}
  {\bibfnamefont {R.~M.}\ \bibnamefont {Geilhufe}},\ and\ \bibinfo {author}
  {\bibfnamefont {H.~B.}\ \bibnamefont {Hamed}},\ }\href
  {https://doi.org/10.1002/pssb.201900671} {\bibfield  {journal} {\bibinfo
  {journal} {physica status solidi (b)}\ }\textbf {\bibinfo {volume} {257}},\
  \bibinfo {pages} {1900671} (\bibinfo {year} {2020})}\BibitemShut {NoStop}%
\bibitem [{\citenamefont {Miranda}\ \emph {et~al.}(1983)\citenamefont
  {Miranda}, \citenamefont {Chandesris},\ and\ \citenamefont
  {Lecante}}]{Miranda1983}%
  \BibitemOpen
  \bibfield  {author} {\bibinfo {author} {\bibfnamefont {R.}~\bibnamefont
  {Miranda}}, \bibinfo {author} {\bibfnamefont {D.}~\bibnamefont
  {Chandesris}},\ and\ \bibinfo {author} {\bibfnamefont {J.}~\bibnamefont
  {Lecante}},\ }\href {https://doi.org/10.1016/0039-6028(83)90361-8} {\bibfield
   {journal} {\bibinfo  {journal} {Surface Science}\ }\textbf {\bibinfo
  {volume} {130}},\ \bibinfo {pages} {269} (\bibinfo {year}
  {1983})}\BibitemShut {NoStop}%
\bibitem [{\citenamefont {S{\'{a}}nchez-Barriga}\ \emph
  {et~al.}(2012)\citenamefont {S{\'{a}}nchez-Barriga}, \citenamefont {Braun},
  \citenamefont {Min{\'{a}}r}, \citenamefont {Marco}, \citenamefont
  {Varykhalov}, \citenamefont {Rader}, \citenamefont {Boni}, \citenamefont
  {Bellini}, \citenamefont {Manghi}, \citenamefont {Ebert}, \citenamefont
  {Katsnelson}, \citenamefont {Lichtenstein}, \citenamefont {Eriksson},
  \citenamefont {Eberhardt}, \citenamefont {Dürr},\ and\ \citenamefont
  {Fink}}]{SanchezBarriga2012}%
  \BibitemOpen
  \bibfield  {author} {\bibinfo {author} {\bibfnamefont {J.}~\bibnamefont
  {S{\'{a}}nchez-Barriga}}, \bibinfo {author} {\bibfnamefont {J.}~\bibnamefont
  {Braun}}, \bibinfo {author} {\bibfnamefont {J.}~\bibnamefont {Min{\'{a}}r}},
  \bibinfo {author} {\bibfnamefont {I.~D.}\ \bibnamefont {Marco}}, \bibinfo
  {author} {\bibfnamefont {A.}~\bibnamefont {Varykhalov}}, \bibinfo {author}
  {\bibfnamefont {O.}~\bibnamefont {Rader}}, \bibinfo {author} {\bibfnamefont
  {V.}~\bibnamefont {Boni}}, \bibinfo {author} {\bibfnamefont {V.}~\bibnamefont
  {Bellini}}, \bibinfo {author} {\bibfnamefont {F.}~\bibnamefont {Manghi}},
  \bibinfo {author} {\bibfnamefont {H.}~\bibnamefont {Ebert}}, \bibinfo
  {author} {\bibfnamefont {M.~I.}\ \bibnamefont {Katsnelson}}, \bibinfo
  {author} {\bibfnamefont {A.~I.}\ \bibnamefont {Lichtenstein}}, \bibinfo
  {author} {\bibfnamefont {O.}~\bibnamefont {Eriksson}}, \bibinfo {author}
  {\bibfnamefont {W.}~\bibnamefont {Eberhardt}}, \bibinfo {author}
  {\bibfnamefont {H.~A.}\ \bibnamefont {Dürr}},\ and\ \bibinfo {author}
  {\bibfnamefont {J.}~\bibnamefont {Fink}},\ }\bibfield  {journal} {\bibinfo
  {journal} {Physical Review B}\ }\textbf {\bibinfo {volume} {85}},\ \href
  {https://doi.org/10.1103/physrevb.85.205109} {10.1103/physrevb.85.205109}
  (\bibinfo {year} {2012})\BibitemShut {NoStop}%
\bibitem [{\citenamefont {Nabok}\ \emph {et~al.}(2021)\citenamefont {Nabok},
  \citenamefont {Blügel},\ and\ \citenamefont {Friedrich}}]{Nabok2021}%
  \BibitemOpen
  \bibfield  {author} {\bibinfo {author} {\bibfnamefont {D.}~\bibnamefont
  {Nabok}}, \bibinfo {author} {\bibfnamefont {S.}~\bibnamefont {Blügel}},\
  and\ \bibinfo {author} {\bibfnamefont {C.}~\bibnamefont {Friedrich}},\
  }\bibfield  {journal} {\bibinfo  {journal} {npj Computational Materials}\
  }\textbf {\bibinfo {volume} {7}},\ \href
  {https://doi.org/10.1038/s41524-021-00649-8} {10.1038/s41524-021-00649-8}
  (\bibinfo {year} {2021})\BibitemShut {NoStop}%
\bibitem [{\citenamefont {M\"uller}\ \emph {et~al.}(2016)\citenamefont
  {M\"uller}, \citenamefont {Friedrich},\ and\ \citenamefont
  {Bl\"ugel}}]{Mueller2016}%
  \BibitemOpen
  \bibfield  {author} {\bibinfo {author} {\bibfnamefont {M.~C. T.~D.}\
  \bibnamefont {M\"uller}}, \bibinfo {author} {\bibfnamefont {C.}~\bibnamefont
  {Friedrich}},\ and\ \bibinfo {author} {\bibfnamefont {S.}~\bibnamefont
  {Bl\"ugel}},\ }\bibfield  {journal} {\bibinfo  {journal} {Physical Review B}\
  }\textbf {\bibinfo {volume} {94}},\ \href
  {https://doi.org/10.1103/physrevb.94.064433} {10.1103/physrevb.94.064433}
  (\bibinfo {year} {2016})\BibitemShut {NoStop}%
\bibitem [{\citenamefont {Müller}\ \emph {et~al.}(2019)\citenamefont
  {Müller}, \citenamefont {Blügel},\ and\ \citenamefont
  {Friedrich}}]{Mueller2019}%
  \BibitemOpen
  \bibfield  {author} {\bibinfo {author} {\bibfnamefont {M.~C. T.~D.}\
  \bibnamefont {Müller}}, \bibinfo {author} {\bibfnamefont {S.}~\bibnamefont
  {Blügel}},\ and\ \bibinfo {author} {\bibfnamefont {C.}~\bibnamefont
  {Friedrich}},\ }\href {https://doi.org/10.1103/physrevb.100.045130}
  {\bibfield  {journal} {\bibinfo  {journal} {Physical Review B}\ }\textbf
  {\bibinfo {volume} {100}},\ \bibinfo {pages} {045130} (\bibinfo {year}
  {2019})}\BibitemShut {NoStop}%
\bibitem [{\citenamefont {Hertz}\ and\ \citenamefont
  {Edwards}(1973)}]{Hertz1973}%
  \BibitemOpen
  \bibfield  {author} {\bibinfo {author} {\bibfnamefont {J.~A.}\ \bibnamefont
  {Hertz}}\ and\ \bibinfo {author} {\bibfnamefont {D.~M.}\ \bibnamefont
  {Edwards}},\ }\href@noop {} {\bibfield  {journal} {\bibinfo  {journal}
  {Journal of Physics F: Metal Physics}\ }\textbf {\bibinfo {volume} {3}},\
  \bibinfo {pages} {2174} (\bibinfo {year} {1973})}\BibitemShut {NoStop}%
\bibitem [{\citenamefont {Edwards}\ and\ \citenamefont
  {Hertz}(1973)}]{Edwards1973}%
  \BibitemOpen
  \bibfield  {author} {\bibinfo {author} {\bibfnamefont {D.~M.}\ \bibnamefont
  {Edwards}}\ and\ \bibinfo {author} {\bibfnamefont {J.~A.}\ \bibnamefont
  {Hertz}},\ }\href@noop {} {\bibfield  {journal} {\bibinfo  {journal} {Journal
  of Physics F: Metal Physics}\ }\textbf {\bibinfo {volume} {3}},\ \bibinfo
  {pages} {2191} (\bibinfo {year} {1973})}\BibitemShut {NoStop}%
\bibitem [{\citenamefont {Hedin}(1965)}]{Hedin1965}%
  \BibitemOpen
  \bibfield  {author} {\bibinfo {author} {\bibfnamefont {L.}~\bibnamefont
  {Hedin}},\ }\href {https://doi.org/10.1103/PhysRev.139.A796} {\bibfield
  {journal} {\bibinfo  {journal} {Phys. Rev.}\ }\textbf {\bibinfo {volume}
  {139}},\ \bibinfo {pages} {A796} (\bibinfo {year} {1965})}\BibitemShut
  {NoStop}%
\bibitem [{\citenamefont {Heinz}\ and\ \citenamefont
  {Hammer}(2009)}]{Heinz2009}%
  \BibitemOpen
  \bibfield  {author} {\bibinfo {author} {\bibfnamefont {K.}~\bibnamefont
  {Heinz}}\ and\ \bibinfo {author} {\bibfnamefont {L.}~\bibnamefont {Hammer}},\
  }\href {https://doi.org/10.1016/j.progsurf.2008.10.003} {\bibfield  {journal}
  {\bibinfo  {journal} {Progress in Surface Science}\ }\textbf {\bibinfo
  {volume} {84}},\ \bibinfo {pages} {2} (\bibinfo {year} {2009})}\BibitemShut
  {NoStop}%
\bibitem [{\citenamefont {Nouvertn{\'{e}}}\ \emph {et~al.}(1999)\citenamefont
  {Nouvertn{\'{e}}}, \citenamefont {May}, \citenamefont {Bamming},
  \citenamefont {Rampe}, \citenamefont {Korte}, \citenamefont {Güntherodt},
  \citenamefont {Pentcheva},\ and\ \citenamefont {Scheffler}}]{Nouvertne1999}%
  \BibitemOpen
  \bibfield  {author} {\bibinfo {author} {\bibfnamefont {F.}~\bibnamefont
  {Nouvertn{\'{e}}}}, \bibinfo {author} {\bibfnamefont {U.}~\bibnamefont
  {May}}, \bibinfo {author} {\bibfnamefont {M.}~\bibnamefont {Bamming}},
  \bibinfo {author} {\bibfnamefont {A.}~\bibnamefont {Rampe}}, \bibinfo
  {author} {\bibfnamefont {U.}~\bibnamefont {Korte}}, \bibinfo {author}
  {\bibfnamefont {G.}~\bibnamefont {Güntherodt}}, \bibinfo {author}
  {\bibfnamefont {R.}~\bibnamefont {Pentcheva}},\ and\ \bibinfo {author}
  {\bibfnamefont {M.}~\bibnamefont {Scheffler}},\ }\href
  {https://doi.org/10.1103/physrevb.60.14382} {\bibfield  {journal} {\bibinfo
  {journal} {Physical Review B}\ }\textbf {\bibinfo {volume} {60}},\ \bibinfo
  {pages} {14382} (\bibinfo {year} {1999})}\BibitemShut {NoStop}%
\bibitem [{sup()}]{supplement}%
  \BibitemOpen
  \href@noop {} {\bibinfo {title} {Supplementary material}}\BibitemShut
  {NoStop}%
\bibitem [{\citenamefont {Liechtenstein}\ \emph {et~al.}(1987)\citenamefont
  {Liechtenstein}, \citenamefont {Katsnelson}, \citenamefont {Antropov},\ and\
  \citenamefont {Gubanova}}]{Liechtenstein1987}%
  \BibitemOpen
  \bibfield  {author} {\bibinfo {author} {\bibfnamefont {A.~I.}\ \bibnamefont
  {Liechtenstein}}, \bibinfo {author} {\bibfnamefont {M.~I.}\ \bibnamefont
  {Katsnelson}}, \bibinfo {author} {\bibfnamefont {V.~P.}\ \bibnamefont
  {Antropov}},\ and\ \bibinfo {author} {\bibfnamefont {V.~A.}\ \bibnamefont
  {Gubanova}},\ }\href {https://doi.org/10.1016/0304-8853(87)90721-9}
  {\bibfield  {journal} {\bibinfo  {journal} {Journal of Magnetism and Magnetic
  Materials}\ }\textbf {\bibinfo {volume} {67}},\ \bibinfo {pages} {65}
  (\bibinfo {year} {1987})}\BibitemShut {NoStop}%
\bibitem [{\citenamefont {Buczek}\ \emph {et~al.}(2009)\citenamefont {Buczek},
  \citenamefont {Ernst}, \citenamefont {Bruno},\ and\ \citenamefont
  {Sandratskii}}]{Buczek2009}%
  \BibitemOpen
  \bibfield  {author} {\bibinfo {author} {\bibfnamefont {P.}~\bibnamefont
  {Buczek}}, \bibinfo {author} {\bibfnamefont {A.}~\bibnamefont {Ernst}},
  \bibinfo {author} {\bibfnamefont {P.}~\bibnamefont {Bruno}},\ and\ \bibinfo
  {author} {\bibfnamefont {L.~M.}\ \bibnamefont {Sandratskii}},\ }\href
  {http://dx.doi.org/10.1103/PhysRevLett.102.247206} {\bibfield  {journal}
  {\bibinfo  {journal} {Phys. Rev. Lett.}\ }\textbf {\bibinfo {volume} {102}},\
  \bibinfo {pages} {247206} (\bibinfo {year} {2009})}\BibitemShut {NoStop}%
\bibitem [{\citenamefont {Schneider}\ \emph {et~al.}(1990)\citenamefont
  {Schneider}, \citenamefont {Bressler}, \citenamefont {Schuster},
  \citenamefont {Kirschner}, \citenamefont {de~Miguel},\ and\ \citenamefont
  {Miranda}}]{Schneider1990}%
  \BibitemOpen
  \bibfield  {author} {\bibinfo {author} {\bibfnamefont {C.~M.}\ \bibnamefont
  {Schneider}}, \bibinfo {author} {\bibfnamefont {P.}~\bibnamefont {Bressler}},
  \bibinfo {author} {\bibfnamefont {P.}~\bibnamefont {Schuster}}, \bibinfo
  {author} {\bibfnamefont {J.}~\bibnamefont {Kirschner}}, \bibinfo {author}
  {\bibfnamefont {J.~J.}\ \bibnamefont {de~Miguel}},\ and\ \bibinfo {author}
  {\bibfnamefont {R.}~\bibnamefont {Miranda}},\ }\href
  {https://doi.org/10.1103/physrevlett.64.1059} {\bibfield  {journal} {\bibinfo
   {journal} {Physical Review Letters}\ }\textbf {\bibinfo {volume} {64}},\
  \bibinfo {pages} {1059} (\bibinfo {year} {1990})}\BibitemShut {NoStop}%
\bibitem [{\citenamefont {Arias}\ and\ \citenamefont
  {Mills}(1999)}]{Arias1999}%
  \BibitemOpen
  \bibfield  {author} {\bibinfo {author} {\bibfnamefont {R.}~\bibnamefont
  {Arias}}\ and\ \bibinfo {author} {\bibfnamefont {D.~L.}\ \bibnamefont
  {Mills}},\ }\href {https://doi.org/10.1103/physrevb.60.7395} {\bibfield
  {journal} {\bibinfo  {journal} {Physical Review B}\ }\textbf {\bibinfo
  {volume} {60}},\ \bibinfo {pages} {7395} (\bibinfo {year}
  {1999})}\BibitemShut {NoStop}%
\bibitem [{\citenamefont {McMichael}\ and\ \citenamefont
  {Krivosik}(2004)}]{McMichael2004}%
  \BibitemOpen
  \bibfield  {author} {\bibinfo {author} {\bibfnamefont {R.}~\bibnamefont
  {McMichael}}\ and\ \bibinfo {author} {\bibfnamefont {P.}~\bibnamefont
  {Krivosik}},\ }\href {https://doi.org/10.1109/tmag.2003.821564} {\bibfield
  {journal} {\bibinfo  {journal} {{IEEE} Transactions on Magnetics}\ }\textbf
  {\bibinfo {volume} {40}},\ \bibinfo {pages} {2} (\bibinfo {year}
  {2004})}\BibitemShut {NoStop}%
\bibitem [{\citenamefont {Zakeri}\ \emph {et~al.}(2007)\citenamefont {Zakeri},
  \citenamefont {Lindner}, \citenamefont {Barsukov}, \citenamefont
  {Meckenstock}, \citenamefont {Farle}, \citenamefont {von Hörsten},
  \citenamefont {Wende}, \citenamefont {Keune}, \citenamefont {Rocker},
  \citenamefont {Kalarickal}, \citenamefont {Lenz}, \citenamefont {Kuch},
  \citenamefont {Baberschke},\ and\ \citenamefont {Frait}}]{Zakeri2007}%
  \BibitemOpen
  \bibfield  {author} {\bibinfo {author} {\bibfnamefont {K.}~\bibnamefont
  {Zakeri}}, \bibinfo {author} {\bibfnamefont {J.}~\bibnamefont {Lindner}},
  \bibinfo {author} {\bibfnamefont {I.}~\bibnamefont {Barsukov}}, \bibinfo
  {author} {\bibfnamefont {R.}~\bibnamefont {Meckenstock}}, \bibinfo {author}
  {\bibfnamefont {M.}~\bibnamefont {Farle}}, \bibinfo {author} {\bibfnamefont
  {U.}~\bibnamefont {von Hörsten}}, \bibinfo {author} {\bibfnamefont
  {H.}~\bibnamefont {Wende}}, \bibinfo {author} {\bibfnamefont
  {W.}~\bibnamefont {Keune}}, \bibinfo {author} {\bibfnamefont
  {J.}~\bibnamefont {Rocker}}, \bibinfo {author} {\bibfnamefont {S.~S.}\
  \bibnamefont {Kalarickal}}, \bibinfo {author} {\bibfnamefont
  {K.}~\bibnamefont {Lenz}}, \bibinfo {author} {\bibfnamefont {W.}~\bibnamefont
  {Kuch}}, \bibinfo {author} {\bibfnamefont {K.}~\bibnamefont {Baberschke}},\
  and\ \bibinfo {author} {\bibfnamefont {Z.}~\bibnamefont {Frait}},\ }\bibfield
   {journal} {\bibinfo  {journal} {Physical Review B}\ }\textbf {\bibinfo
  {volume} {76}},\ \href {https://doi.org/10.1103/physrevb.76.104416}
  {10.1103/physrevb.76.104416} (\bibinfo {year} {2007})\BibitemShut {NoStop}%
\bibitem [{\citenamefont {Buczek}\ \emph {et~al.}(2020)\citenamefont {Buczek},
  \citenamefont {Buczek}, \citenamefont {Vignale},\ and\ \citenamefont
  {Ernst}}]{Buczek2020}%
  \BibitemOpen
  \bibfield  {author} {\bibinfo {author} {\bibfnamefont {P.}~\bibnamefont
  {Buczek}}, \bibinfo {author} {\bibfnamefont {N.}~\bibnamefont {Buczek}},
  \bibinfo {author} {\bibfnamefont {G.}~\bibnamefont {Vignale}},\ and\ \bibinfo
  {author} {\bibfnamefont {A.}~\bibnamefont {Ernst}},\ }\href
  {https://doi.org/10.1103/physrevb.101.214420} {\bibfield  {journal} {\bibinfo
   {journal} {Phys. Rev. B}\ }\textbf {\bibinfo {volume} {101}},\ \bibinfo
  {pages} {214420} (\bibinfo {year} {2020})}\BibitemShut {NoStop}%
\end{thebibliography}%

\end{document}